\documentclass[12pt,a4paper]{iopart}
\usepackage{iopams}
\usepackage[english]{babel}
\usepackage{float}
\usepackage[dvips]{graphicx}
\usepackage{rotating}

\usepackage{epsfig}
\usepackage[dvips]{color}
\definecolor{Red}{named}{Red}

\begin{document}
\title{Grid Based Linear Neutrino Perturbations in Cosmological $N$-body Simulations}
\author{Jacob Brandbyge$^1$, Steen Hannestad$^1$}
\ead{jacobb@phys.au.dk, sth@phys.au.dk}
\address{$^1$Department of Physics and Astronomy, University
of Aarhus, Ny Munkegade, DK-8000 Aarhus C, Denmark}
\date{\today}

\begin{abstract}
We present a novel, fast and precise method for including the effect of light neutrinos in cosmological $N$-body simulations. The effect of the neutrino component is included by using the linear theory neutrino perturbations in the calculation of the gravitational potential in the $N$-body simulation. By comparing this new method with the full non-linear evolution first presented in \cite{Brandbyge1}, where the neutrino component was treated as particles, we find that the new method calculates the matter power spectrum with an accuracy better than 1\% for $\sum m_\nu \lesssim 0.5 \, {\rm eV}$ at $z = 0$. This error scales approximately as $(\sum m_\nu)^2$, making the new linear neutrino method extremely accurate for a total neutrino mass in the range $0.05 - 0.3 \, {\rm eV}$. At $z = 1$ the error is below 0.3\% for $\sum m_\nu \lesssim 0.5 \, {\rm eV}$ and becomes negligible at higher redshifts. This new method is computationally much more efficient than representing the neutrino component by $N$-body particles.
\end{abstract}
\maketitle

\section{Introduction}

Next to photons neutrinos are the most abundant particles in our Universe. The fact that at least two of the three neutrino mass eigenstates have masses much larger than the current temperature, $m_i \gg T_0$, means that neutrinos contribute to the matter density and are important for cosmological structure formation.
The mass differences established by oscillation experiments,
$\Delta{m}^2_{12}\simeq 7.6 \times
10^{-5} \, {\rm eV}^2$ and $|\Delta{m}^2_{23}| \simeq 2.4 \times 10^{-3} \, {\rm eV}^2$ (see e.g.\ \cite{Schwetz:2008er} for a recent
data analysis), mean that the heaviest mass eigenstate must have a mass of at least $m \sim 0.05$ eV.
This in turn will have the effect of suppressing the power spectrum of matter fluctuations by $\sim 5$\%, an effect which is significantly larger than the precision with which the matter power spectrum can be measured in upcoming surveys.

This has two interesting consequences, first it means that neutrino mass {\it must} be included in the analysis of future data in order to avoid seriously biasing the measurements of parameters such as the dark energy equation of state, $w$ \cite{Hannestad:2005gj}. Second, it means that future large-scale structure surveys potentially have the power to probe neutrino masses as low as the current {\it lower} bound from oscillation experiments \cite{futuremass}.

Consequently this has led to a much increased interest in gaining a detailed understanding of how neutrinos affect structure formation. In linear theory the effect is extremely well understood and can be calculated to a precision much better than 1\%. However, these results apply only on very large scales, $k \ll 0.1 \, h {\rm Mpc}^{-1}$, whereas most of the cosmologically relevant information from large-scale structure surveys is in the range $k \sim 0.1-0.5 \, h {\rm Mpc}^{-1}$.
On these intermediate scales it is mandatory to correct for non-linear effects, even if surveys are carried out at intermediate redshifts where non-linearity is weaker.

Non-linear corrections can be found in a number of ways: The most accurate, but also most time-consuming method is to use full $N$-body simulations with neutrinos included in a self-consistent way. In a previous publication we carried out a large suite of simulations with neutrinos included \cite{Brandbyge1}, and found a significant non-linear correction caused by neutrinos. The effect is large enough to be important even for masses as low as 0.05 - 0.1 eV.
Another way is to use higher order perturbation theory which, however, is applicable only up to $k \sim 0.1-0.2 \, h {\rm Mpc}^{-1}$ today, if the precision needs to be better than 1-2\%. Finally it is possible to use semi-analytic methods such as the halo model which has been calibrated using $\Lambda$CDM simulations. Neutrinos can be included fairly easily in this scheme, but the accuracy can only be tested against full $N$-body simulations.

In conclusion, a very large number of high-resolution $N$-body simulations with neutrinos included will be necessary for future data analysis.
The method used in \cite{Brandbyge1} is well suited for large neutrino masses with $\sum m_\nu \gtrsim 0.5$ eV at low redshift, but for all masses the simulations inevitably become noise dominated at high redshift, unless extremely many neutrino $N$-body particles are used.

In the present paper we describe a novel, fast and precise method for simulating low mass neutrinos in $N$-body simulations which uses the fact that low mass neutrinos have very little non-linear clustering even at low redshift.
Instead of simulating neutrinos as particles with individual velocities we describe the local neutrino density on a grid. This density is evolved forward in time by using linear theory. The CDM/baryon component is followed simultaneously using the full TreePM method, with the neutrino component contributing to the long range force calculated using the PM method.
As will be shown below the accuracy of this method is better than 1\% for the matter power spectrum today on {\it all} scales for all neutrino masses below $\sum m_\nu \sim 0.5$ eV.

The method is very powerful, both because it speeds up simulations with neutrinos very significantly (by factors of order $\sim 10$), {\it and} because it can be applied to any other component of energy density which is almost linear. This could for example be dark energy with linear perturbations. The only requirement is that the linear component can be included in a Boltzmann solver like CAMB \cite{CAMB}.

We note that our new method is conceptually similar to the perturbative approaches presented in \cite{saito1, Wong08}, except for the fact that we give the CDM component a full non-linear treatment.

In Section \ref{IC} we describe our cosmological model and the set-up of initial conditions for our neutrino {\it N}-body
simulations. In Section \ref{RESULTS} we present our results and in Section \ref{CONVERGE} we show that our results have converged. Finally, Section \ref{DISCUSSION} contains a discussion and our conclusions.

\section{Initial Conditions}\label{IC}
\subsection{The Cosmological Model and Particle Initial Conditions}

As long as the evolved primordial density perturbations set down by inflation remain small, their evolution can be calculated precisely using the
linearised Einstein and Boltzmann equations \cite{Ma}. However, once structure enters the non-linear regime precision
studies require the use of {\it N}-body simulations. To set up the initial conditions (ICs) for our simulations we have calculated the
transfer functions (TFs) using CAMB \cite{CAMB}. The further evolution is followed in \textsc{gadget}-2 \cite{Springel2} run in the hybrid TreePM mode.

We have assumed a flat cosmological model with density parameters $\Omega_{\rm b}=0.05$,
$\Omega_{\rm m}=0.30$ and $\Omega_\Lambda=0.70$ for the baryon, total matter and
cosmological constant components, respectively, and a Hubble parameter
of $h = 0.70$. We vary the CDM and neutrino density parameters
($\Omega_{\rm CDM}$ and $\Omega_\nu$, respectively) such that they fullfill the
condition $\Omega_{\rm CDM}+\Omega_\nu = 0.25$. We have assumed a primordial power spectrum of the standard scale-invariant Harrison-Zel'dovich form. The amplitude gives $\sigma_8=0.878$ for a pure $\Lambda {\rm CDM}$ model.

To generate the ICs for our simulations we have built a parallelized IC generator directly into \textsc{gadget}-2. The TFs are used to generate the position and velocity ICs for the $N$-body particles. In addition to the Zel'dovich
Approximation (ZA) \cite{Zeldovich:1969sb}, we have included a correction term from second-order Lagrangian perturbation theory (2LPT)
\cite{Scoccimarro1,Bouchet}. Due to our high $N$-body starting redshift ($z = 49$), this second-order term is small, and therefore basically does not affect our results.

The neutrino component is included in the $N$-body simulation either as real space particles or on a grid in Fourier space. The CDM component is always treated as particles and for its initial power spectrum we have used a weighted sum of the CDM and baryon TFs. Gas physics is not included in the $N$-body simulation, since it does not significantly affect the scales simulated. The initial conditions for the CDM and neutrino components have been generated with the same set of random numbers, to enforce the assumption of adiabatic initial fluctuations. In the neutrino particle method we add a thermal velocity drawn from a relativistic Fermi-Dirac distribution to the neutrino $N$-body particles. When considering the power spectrum this thermal velocity acts as a wavenumber dependent suppression term (see \cite{Brandbyge1} for further details).

\subsection{Initial Conditions and Evolution of the Neutrino Grid}
When the linear neutrino component is represented on a grid in the $N$-body simulation, there are no neutrino $N$-body particles. The representation is done as follows. A realisation of the neutrino TF is generated with the ZA only, using the same set of random numbers as was used to make the initial conditions for the CDM particles. The neutrino component stays in the Fourier domain throughout the $N$-body simulation. The linear neutrino modes are added directly to their CDM counterparts whenever the long range force is calculated, taking proper care of the fact that only the latter modes should be deconvolved from the smoothing effect of the Cloud-in-Cell (CIC) mass assignment used in \textsc{gadget}-2. Whenever the long range force is calculated the neutrino component is evolved by interpolating between a library of TFs which span the redshift range over which the power spectrum is to be evolved in the $N$-body simulation.

The neutrinos are not included in the short range Tree part. To compensate for this fact the neutrino modes are not smoothed in Fourier space with the Gaussian factor ${\rm exp}(-k^2r_s^2)$, where $r_s = 1.25 \, h^{-1} {\rm Mpc}$, as are the CDM modes (see \cite{Bagla02} for an analysis of the TreePM method). This lack of smoothing could be a problem since the accuracy of the PM method breaks down as the PM grid mesh size is approached (see Section \ref{CONVERGE}).

A concern about the neutrino grid method could be that there are rotations in the CDM density field in real space, which could bring the CDM and neutrino components out of phase. But as long as only the linear neutrino density distribution contributes to the matter power spectrum these offsets due to rotations are negligible. Another more serious problem could arise since the linear part of the neutrino power spectrum and the CDM power spectrum are evolved in two different time integrators, CAMB and \textsc{gadget}-2, respectively. Force calculation errors in the two integrators could lead to discrepancies, which would not only affect the absolute power spectrum but also the relative power spectrum between the neutrino grid and particle approaches. In Section \ref{CONVERGE} we address these issues.

\begin{table*}[t]
{\footnotesize 
  \hspace*{-0.8cm}\begin{tabular}
  {|c||c|c@{\hspace{2pt}}c|c@{\hspace{2pt}}c@{\hspace{2pt}}c@{\hspace{2pt}}c@{\hspace{2pt}}c@{\hspace{2pt}}c@{\hspace{2pt}}c@{\hspace{2pt}}c@{\hspace{2pt}}c@{\hspace{2pt}}c|c@{\hspace{2pt}}c@{\hspace{2pt}}c|}                      \hline
  & $A_1$& $B_1$ & $B_2$ & $C_1$ & $C_2$ &$C_3$&$C_4$&$C_5$ & $C_6$ &$C_7$&$C_8$&$C_9$ & $C_{10}$& $D_1$ & $D_2$& $D_3$ \\       \hline\hline
$N_{\rm CDM}$ &$512^3$& $512^3$ & $512^3$ & $512^3$&$512^3$&$512^3$&$256^3$& $256^3$&$256^3$&$256^3$&$512^3$&$512^3$&$512^3$&$512^3$&$512^3$&$512^3$\\
$N_{\nu,{\rm part}}$ &0    & $0$& $256^3$ & 0&$256^3$&$512^3$&$0$&$256^3$ & $0$&$256^3$& $0$& $256^3$ & $0$ & $0$ & $256^3$ & $512^3$\\
$N_{\nu,{\rm grid}}$ &0    & $512^3$ & $0$& $512^3$&0&$0$&$256^3$&$0$   & $512^3$& $0$ & $1024^3$  & 0 &0 & $512^3$ &  $0$ & $0$\\
$N_{\rm PM}$       & 512& 512&512& 512&512&512&256&256&512&512&1024&1024 &512& 512&512 & 512\\
$R_{\rm BOX}$ $[h^{-1} {\rm Mpc}]$  &512&512&512&512&512&512&256&256&512&512&512&512&512 & 512& 512& 512\\
$\sum m_\nu~[{\rm eV}]$ & 0& 0.3&0.3&0.6&0.6&0.6&0.6&0.6&0.6&0.6&0.6&0.6&0.6 &1.2 &1.2& 1.2  \\
$\Omega_{\nu,0}$~[$\%$] & 0& 0.65&0.65& 1.3&1.3&1.3&1.3&1.3&1.3&1.3&1.3&1.3 &1.3& 2.6&2.6          & 2.6\\\hline
  \end{tabular}
  }
  \caption{Parameters for the $N$-body simulations used to make the power spectra presented in this paper. $N_{\rm CDM}$ and $N_{\nu,{\rm part}}$ is the number of CDM and neutrino $N$-body particles, respectively. $N_{\nu,{\rm grid}}$ is the size of the neutrino Fourier grid. $\sum m_\nu $ is the total neutrino mass, and it is in all cases related to the one-particle neutrino mass, $m_\nu$, by $\sum m_\nu = 3 m_\nu$. $\Omega_{\nu,0}$ is the
  fraction of the critical density contributed by the neutrinos today. $N_{\rm PM}$ is the number of one-dimensional PM grid points and $R_{\rm BOX}$ is the size of the simulation volume. All simulations have a starting redshift of 49. Note that in model $C_{10}$ the neutrino distribution has been kept totally homogeneous in the $N$-body simulation. We will use the notation $X_i/X_j$ to indicate a relative power spectrum calculated between the models $X_i$ and $X_j$ from $(X_i / X_j - 1)\cdot 100\%$.}
  \label{fig:table11}
\end{table*}

\section{Results}\label{RESULTS}
\begin{figure}[t]
   \noindent
      \includegraphics[width=0.99\textwidth]{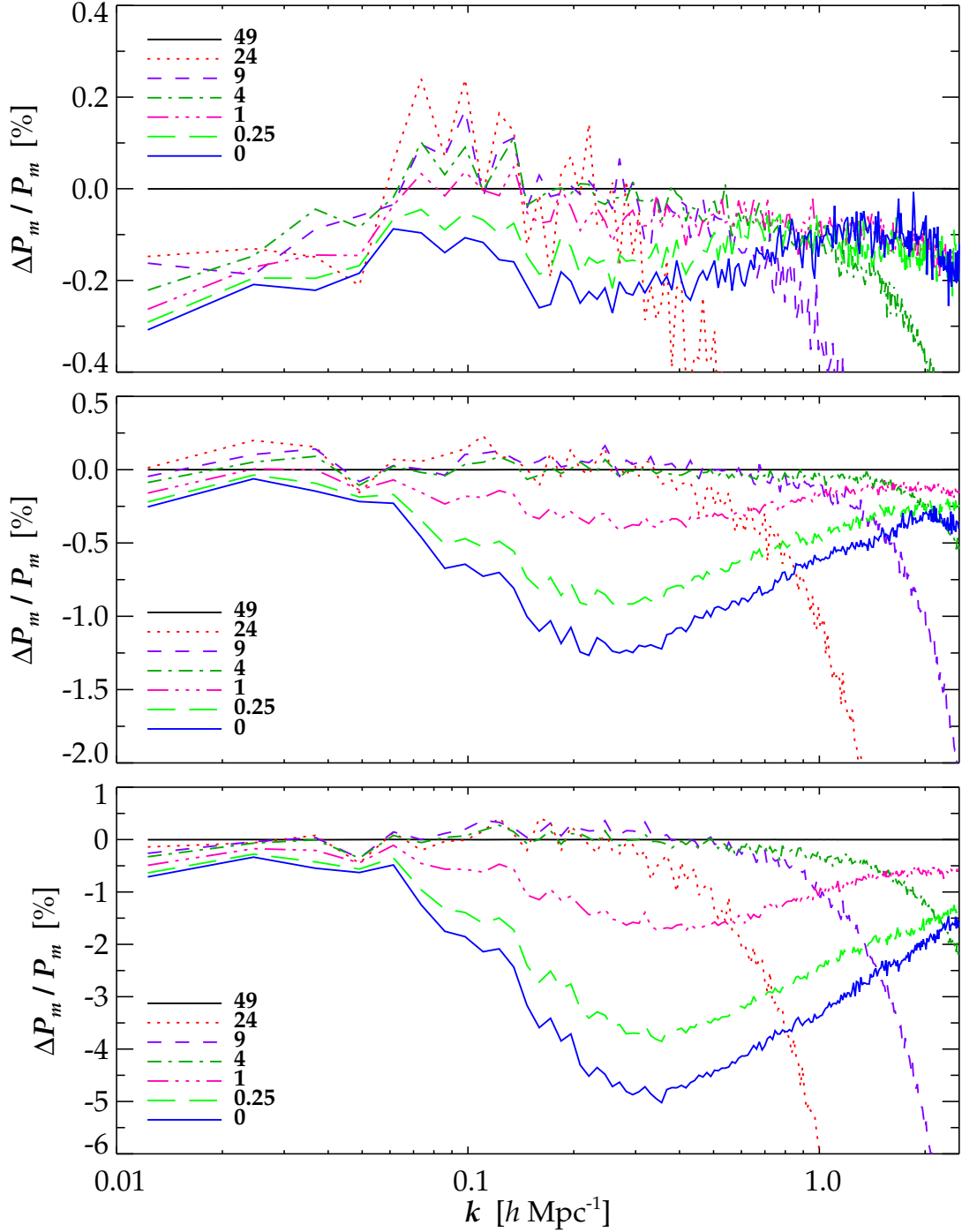}
      \vspace*{1.0cm}
   \caption{Percentage differences in the total matter power spectrum at different redshifts between simulations where the neutrino component is represented in the $N$-body simulation either on a grid or as particles. A negative power difference indicates more power in the particle simulations. From top to bottom the total neutrino mass is $0.3 \, {\rm eV}$ ($B_1 / B_2$), $0.6 \, {\rm eV}$ ($C_1 / C_3$) and $1.2 \, {\rm eV}$ ($D_1 / D_3$), respectively. All the power spectra presented in this paper have been calculated on a $1024^3$ grid, using a deconvolved CIC mass assignment scheme for the $N$-body particles.}
   \label{fig:grid_part_diff}
\end{figure}

\begin{figure}
   \noindent
   \includegraphics[width=1.05\linewidth]{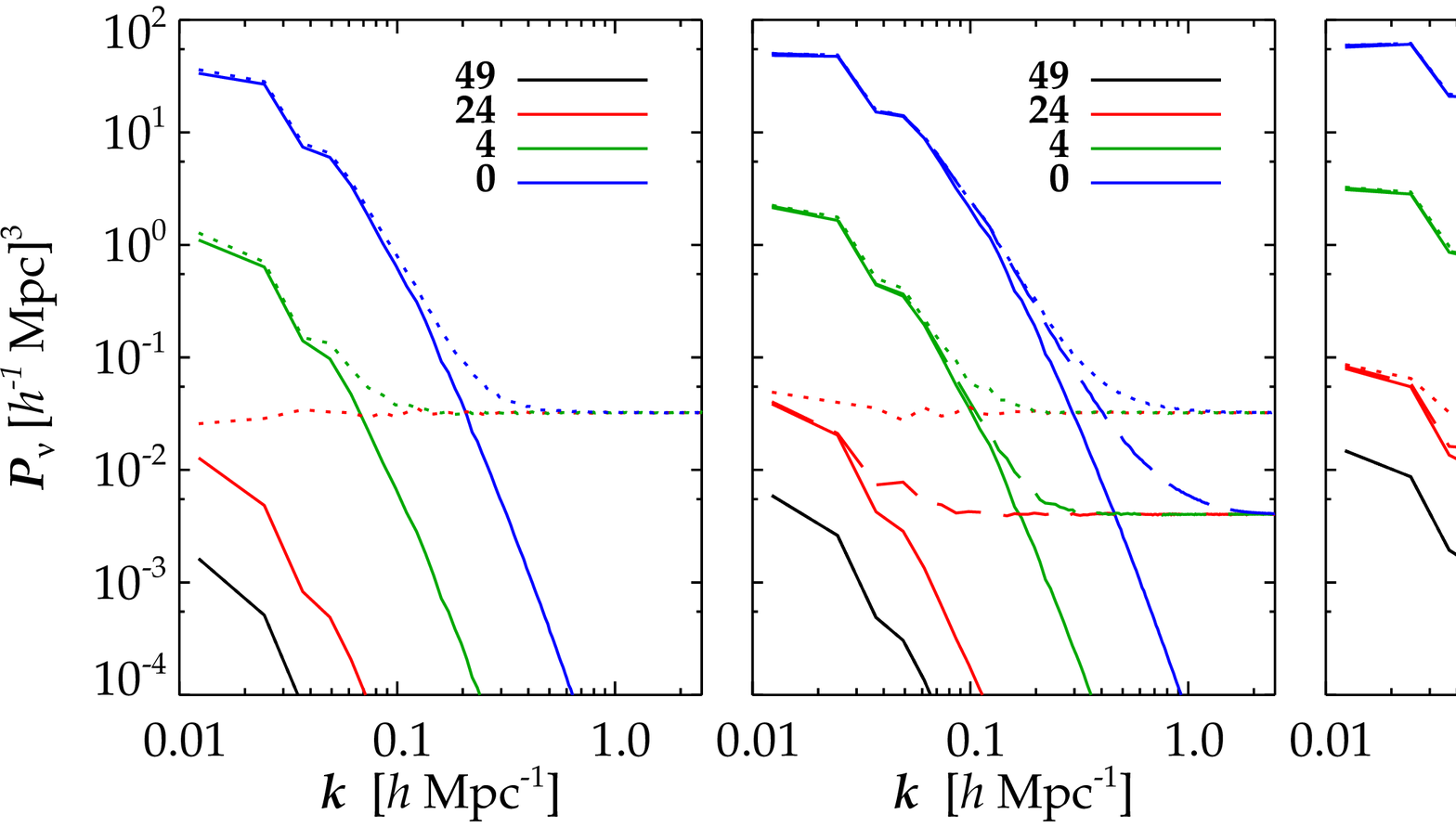}
   \caption{Neutrino power spectra for a total neutrino mass of $0.3 \, {\rm eV}$ (left), $0.6 \, {\rm eV}$ (middle) and $1.2 \, {\rm eV}$ (right) at various redshifts. The linear theory neutrino power spectrum, convolved with our chosen random numbers, are shown with solid lines (models $B_1$, $C_1$ and $D_1$), simulations with $256^3$ neutrino particles with dotted lines (models $B_2$, $C_2$ and $D_2$) and finally simulations with $512^3$ neutrino particles are shown with dashed lines (models $C_3$ and $D_3$).}
   \label{fig:nu_power}
\end{figure}

This section will focus on our main results, whereas the next section will contain a description of detailed convergence tests. Table \ref{fig:table11} shows most of our performed $N$-body simulations.

\begin{figure}[t]
   \includegraphics[width=1.0\linewidth]{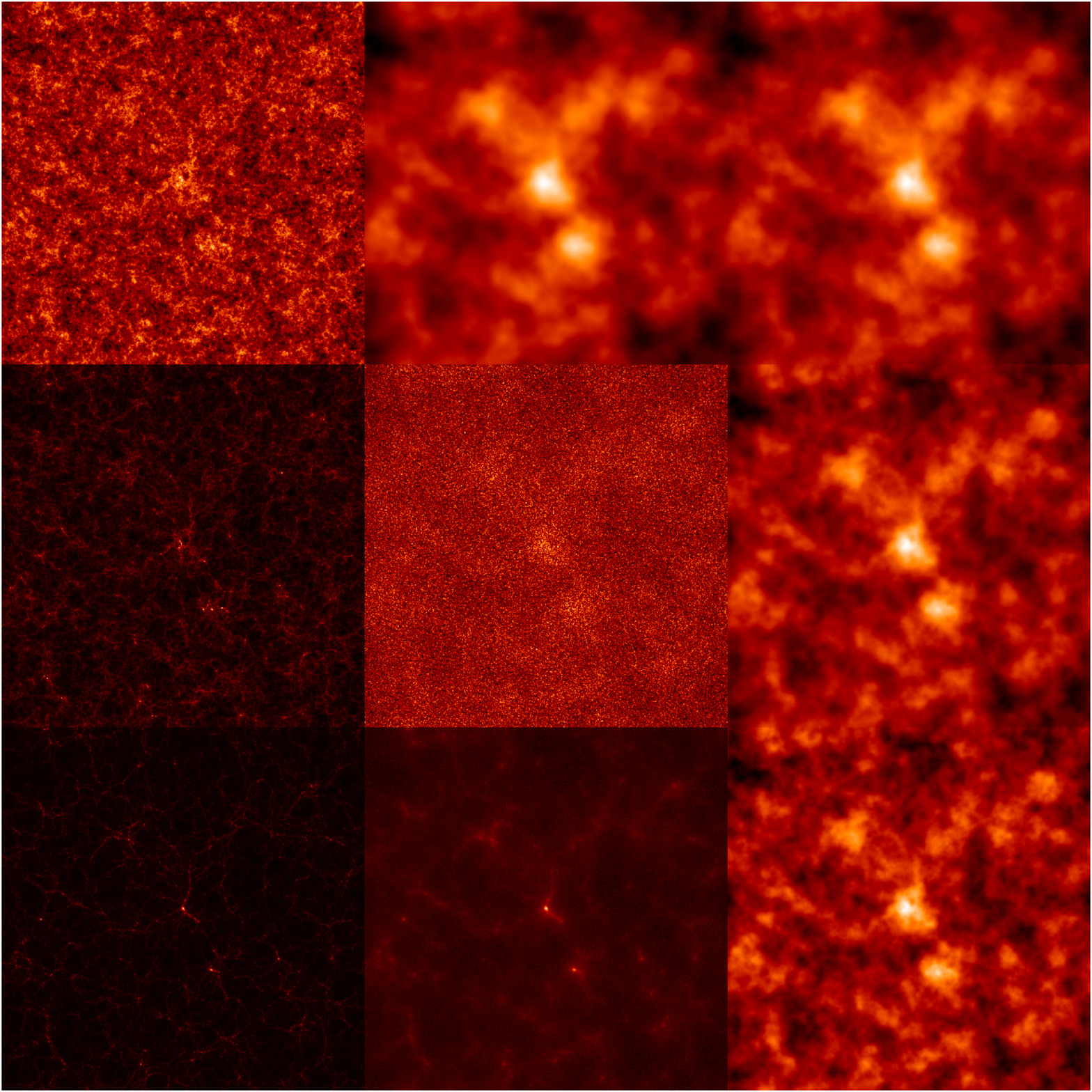}
   \caption{Density grids for the CDM (left), neutrino particle (middle) and neutrino grid (right) components. In all cases the total neutrino mass is $0.6 \, {\rm eV}$. The top row is at $z = 49$, the middle row at $z = 4$, and the bottom row at $z = 0$. In the bottom row the square root has been taken of the first two density distributions. The images are centered at the highest density region in the simulation volume and they have a thickness of $20 \, h^{-1} {\rm Mpc}$ and a side length of $512 \, h^{-1} {\rm Mpc}$. The particle density distributions are found using the adaptive smoothing length kernel from \cite{monaghan} (taken from model $C_3$), and the neutrino grid density distribution is an inverse FFT of the linear neutrino Fourier grid imbedded in the $N$-body simulation volume (model $C_1$).}
   \label{fig:m06rho}
\end{figure}

Fig.~\ref{fig:grid_part_diff} shows the difference in the total matter (neutrino plus CDM) power spectrum for 3 different neutrino masses, between simulations where the neutrino component is represented either on a grid or as $N$-body particles. The difference between the two methods are shown at various redshifts. At our initial $N$-body starting redshift, $z = 49$, the two methods are identical, as they should be. In the simulation where the neutrino component is represented by particles the neutrino $N$-body particles free-stream out of the gravitational potential wells and generate a white noise term as soon as the simulation is started.
This white noise term affects the difference power spectra in Fig.~\ref{fig:grid_part_diff} for $k > 0.1 - 1 h \, {\rm Mpc}^{-1}$ (the actual wavenumber depending on redshift, neutrino mass and the number density of neutrino $N$-body particles), and is solely due to the finite number of neutrino $N$-body particles. That this is indeed a white noise term can be seen from Fig.~\ref{fig:nu_power} where the neutrino power spectrum is shown for all neutrino masses simulated. This white noise term can furthermore be seen in Fig.~\ref{fig:m06rho}, where the neutrino $N$-body particle density distribution is shown at $z = 4$ (i.e.\ the central panel).

From Fig.~\ref{fig:nu_power} it can be seen that the neutrino $N$-body particle white noise term is frozen on small scales, i.e. there is a maximum white noise level, which is determined by the number of neutrino $N$-body particles. Note that for $\sum m_\nu = 0.3 \, {\rm eV}$ and $z = 24$ even the neutrino fundamental mode is white noise dominated. This noise term persists and can be identified in the top panel of Fig.~\ref{fig:grid_part_diff}.

As the simulation evolves the CDM density perturbations grow so that the neutrino white noise term in the matter power spectrum dominates only on ever smaller scales. This is clearly seen in Fig.~\ref{fig:grid_part_diff} where the agreement between the grid and particle representations improves on small scales as the simulation evolves. At $z = 4$ the two methods give basically identical results until $k > 1 h \, {\rm Mpc}^{-1}$. The disagreement beyond this wavenumber is caused by the white noise term.

\begin{figure}[t]
   \hspace*{-0.8cm}\includegraphics[width=1.1\linewidth]{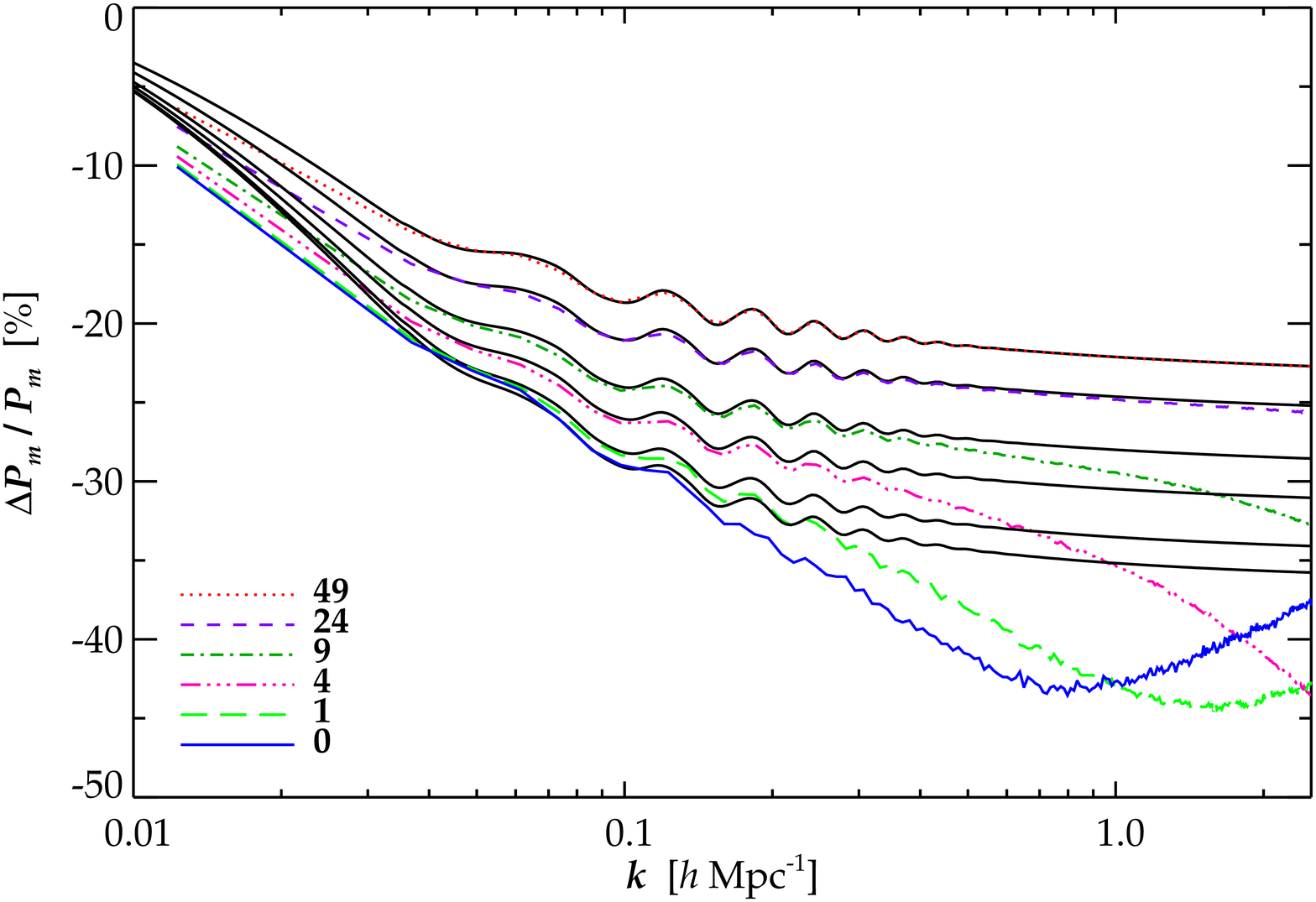}
   \caption{Evolution of the difference in the total matter power spectrum between a pure $\Lambda {\rm CDM}$ model ($A_1$) and a model with $\sum m_\nu = 0.6 \, {\rm eV}$ neutrinos on a grid ($C_1$). The difference expected from linear theory is also shown (black solid lines).}
   \label{fig:bump_evol}
\end{figure}

In sum, for $\sum m_\nu = 0.6 \, {\rm eV}$ the neutrino grid method is accurate at the 0.1\% level for all scales until $z = 4$. This very low discrepancy is impressive given the fact that the neutrino component contribute up to more than 10\% at this redshift (see the left panel of Fig.~\ref{fig:no_nus}), as well as the fact that the neutrino component is evolved in two different integrators.

As the redshift falls below 4 the two representations begin to differ in the range $k \simeq 0.1 - 1 \, h {\rm Mpc}^{-1}$. This difference is a non-linear correction coming from the fact that the neutrino $N$-body particles are coupled to the non-linear gravitational potential whereas the neutrino grid is only evolved in linear theory. The wavenumber range where this difference appears can be explained by the convolution of two terms. A non-linear term growing rapidly on small scales and the fact that the neutrinos contribute most on large scales, as can be seen in Fig.~\ref{fig:no_nus}.

The lowest wavenumber at which this non-linear correction term becomes important, can also be identified in the middle and right panel of Fig.~\ref{fig:nu_power}, as the range from where the particle neutrino power spectra break away from the linear theory evolution. The highest wavenumber at which the non-linear correction term matters is not only determined by non-linear neutrino modes at that scale but also by mode-coupling between the CDM perturbations at that scale and the extra non-linear neutrino contribution at larger scales.

In the $\sum m_\nu = 0.6 \, {\rm eV}$ case at $z = 0$ the two methods differ by at most 1.25\% at roughly $k \simeq 0.25 \, h {\rm Mpc}^{-1}$. Focusing on the $\sum m_\nu = 1.2 \, {\rm eV}$ case\footnote{This high neutrino mass was only used to illustrate when the linear neutrino approach breaks down.}, we see from Fig.~\ref{fig:grid_part_diff} that the non-linear correction is at the 5\% level. As expected the non-linear correction is greater as the neutrino mass is increased because $\Omega_\nu$ increases and the neutrino thermal velocity decreases. Since $\Omega_\nu$ is proportional to the total neutrino mass and the thermal velocity is roughly inverse proportional to the one-particle neutrino mass this explains the $(\sum m_\nu)^2$ dependence on the size of the non-linear neutrino correction term. This scaling can also be seen to roughly hold in the $\sum m_\nu = 0.3 \, {\rm eV}$ case, where the maximum non-linear neutrino correction to the matter power spectrum is at a negligible 0.25\% level today (the discrepancy at large scales is due to a finite number of neutrino $N$-body particles, see Section \ref{CONVERGE}). Note that as expected the wavenumber corresponding to the maximum non-linear neutrino correction propagates slighly from $k \simeq 0.3 \, h{\rm Mpc}^{-1}$ in the $\sum m_\nu = 1.2 \, {\rm eV}$ case to $k \simeq 0.2 \, h {\rm Mpc}^{-1}$ in the $\sum m_\nu = 0.3 \, {\rm eV}$ case. In sum, the percentage maximum non-linear correction to the neutrino component today is well fitted by the relation $({\sum m_\nu}/{0.54 \, {\rm eV}})^2$.

Finally, in Fig.~\ref{fig:bump_evol} we show the evolution of the difference in the matter power spectrum between a pure $\Lambda$CDM simulation ($A_1$) and model $C_1$ with $0.6 \, {\rm eV}$ neutrinos represented on a grid. For the scales simulated it can be seen that the turn-over in the difference power spectrum is created at low redshift, $z \simeq 1 -4$, and that it propagates to larger scales. The almost perfect agreement between linear and non-linear theory seen in the figure at $z = 24$ can in practice only be achieved by representing the neutrino component on a grid.

\section{Convergence Tests}\label{CONVERGE}
In the two approaches the neutrino component is evolved in different integrators. In the grid approach the neutrino component is evolved in CAMB with the full general relativistic equations and no cosmic variance. In the particle approach the neutrino component is evolved in \textsc{gadget}-2, with an accuracy limited by the finite box size, a finite number of CDM and neutrino $N$-body particles (particle shot noise), as well as finite time steps and force resolution. It is important that the neutrino component is evolved accurately in both integrators\footnote{The accuracy with which CAMB calculates the linear power spectrum is far better than 1\% on all scales.}, so that our results are not compromised.

Since we are interested in quantifying the non-linear neutrino correction term, we need to verify that the relative differences in the power spectrum between the two approaches have converged sufficiently. This demand is not as challenging as simulating a converged absolute power spectrum. We note that small-scale differences in the total matter power spectrum are acceptable since this part is dominated by the CDM component, which is evolved in the same integrator in both the grid and particles approaches. Small inaccuracies at small scales will therefore cancel out when comparing power spectra which only differs at the few percent level. But since the two neutrino representations are evolved in different integrators, we need a converged absolute neutrino power spectrum at the scales where the neutrino component contributes to the total matter power spectrum.

Those test runs described in this section which are not listed in Fig.~\ref{fig:table11} are evolved in a $512 \, h^{-1} {\rm Mpc}$ box and all have $256^3$ CDM particles and a $512^3$ PM grid.

\subsection{Initial Velocities and their Evolution}
We have found the gravitationally induced initial $N$-body velocities by taking the time difference between two displacement grids centered around our $N$-body starting redshift by some small $\Delta z$. We have run 3 simulations varying $\Delta z$ in the range $0.5-2$, and found that our choose of $\Delta z \simeq 1$ has converged at a few tens of a percent.

\begin{figure}
   \noindent
   \hspace*{-1.1cm}\includegraphics[width=1.1\linewidth]{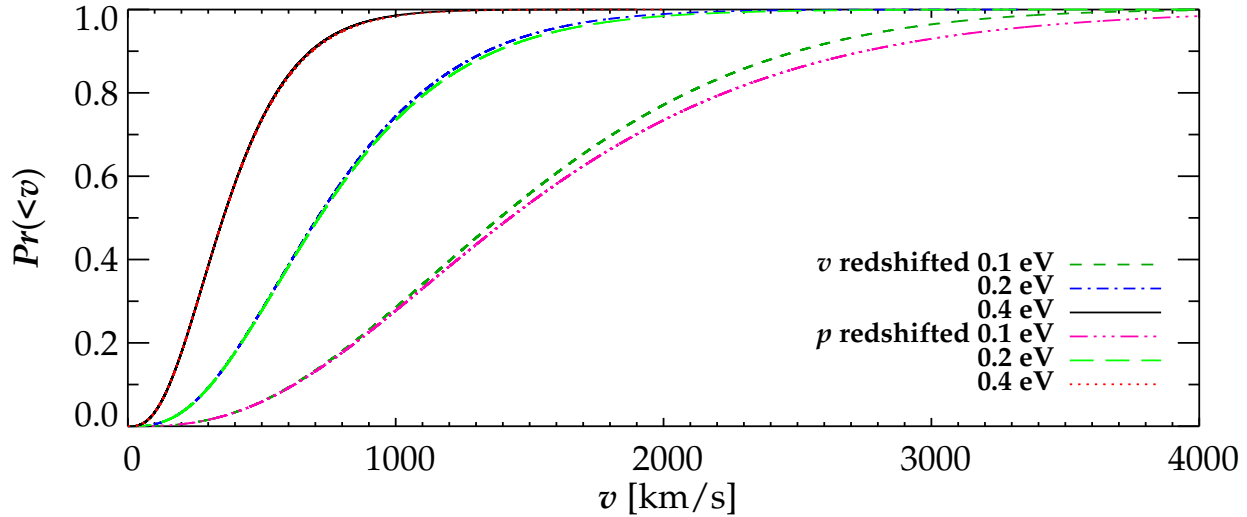}
   \caption{The figure shows the effect of redshifting either the Fermi-Dirac thermal velocity or momentum from $z = 49$ until $z = 0$, for 3 different neutrino one-particle masses.}
   \label{fig:FermiDirac}
\end{figure}

In \textsc{gadget}-2 particle velocities are redshifted, not the momentum. This Newtonian evolution is very well justified for CDM particles but could pose a problem for low mass neutrinos, since at high redshift their Fermi-Dirac distributed thermal velocities can approach half the speed of light. In Fig.~\ref{fig:FermiDirac} we have calculated the cumulative neutrino thermal velocity distribution today in two different ways for our 3 neutrino one-particle masses simulated. Either by redshifting the velocity or the momentum from our $N$-body starting redshift of 49. From the figure it can be seen that for our highest simulated neutrino mass of $0.4 \, {\rm eV}$, it is very accurate to redshift the velocity. In the $0.2 \, {\rm eV}$ case redshifting the velocity or momentum gives the same result except for a slight difference at the high end of the velocity distribution tail. This small difference will not affect our results. In the $0.1 \, {\rm eV}$ case this difference is slightly larger, but since it is mainly the neutrino $N$-body particles with a thermal velocity drawn from the low end of the distribution which contribute to the matter power spectrum, redshifing the velocity is accurate enough. Likewise relativistic velocity addition is not necessary. Note that redshifting the velocity instead of the momentum leads to a lower thermal velocity and therefore $more$ structure formation, so that from this point of view the size of the non-linear correction to the neutrino component in Fig.~\ref{fig:grid_part_diff} is an upper bound.

\begin{figure}[t]
   \noindent
   \begin{minipage}{0.49\linewidth}
      \hspace*{-1.4cm}\includegraphics[width=1.25\linewidth]{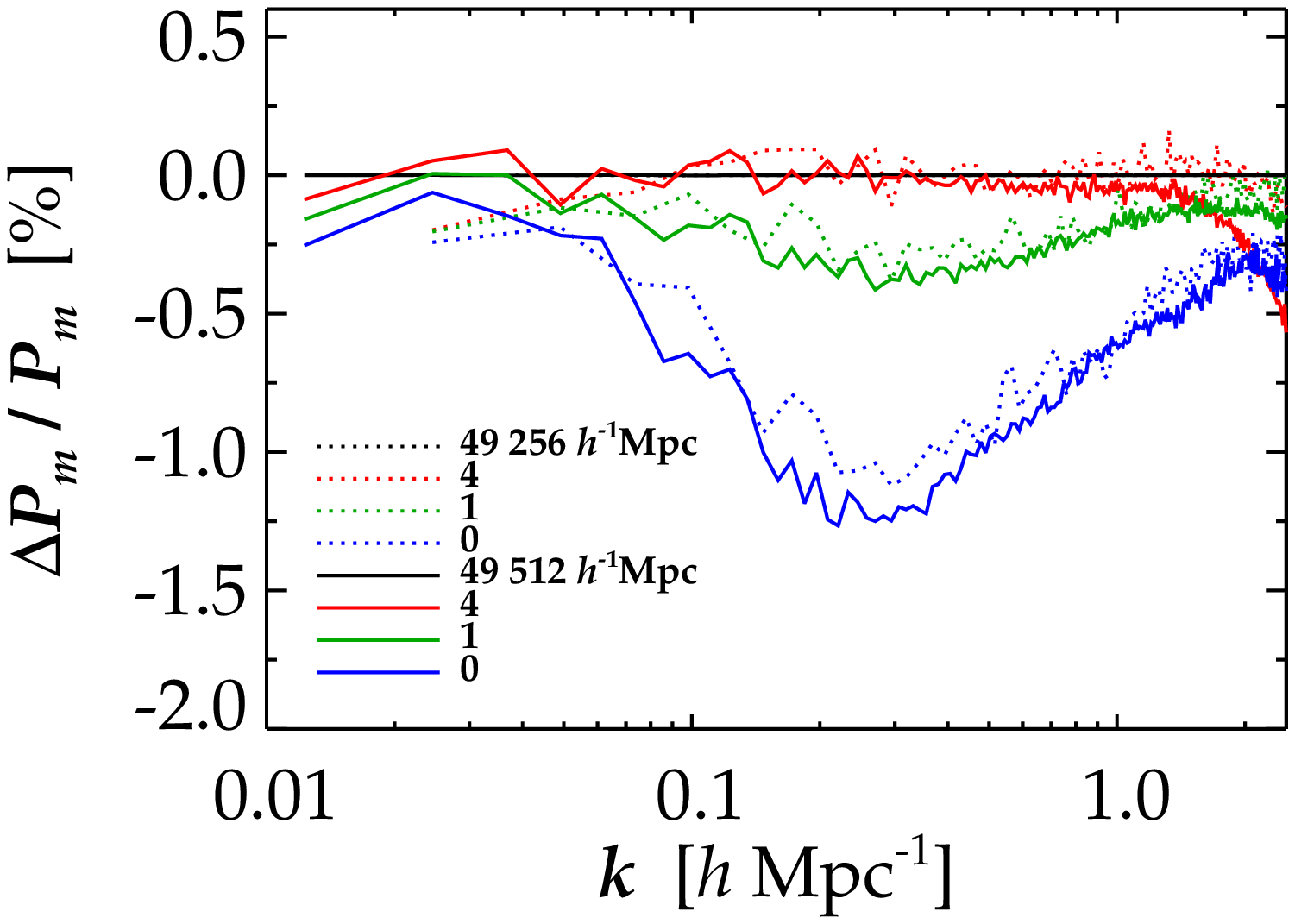}
   \end{minipage}
   \begin{minipage}{0.49\linewidth}
      \hspace*{-0.3cm}\includegraphics[width=1.25\linewidth]{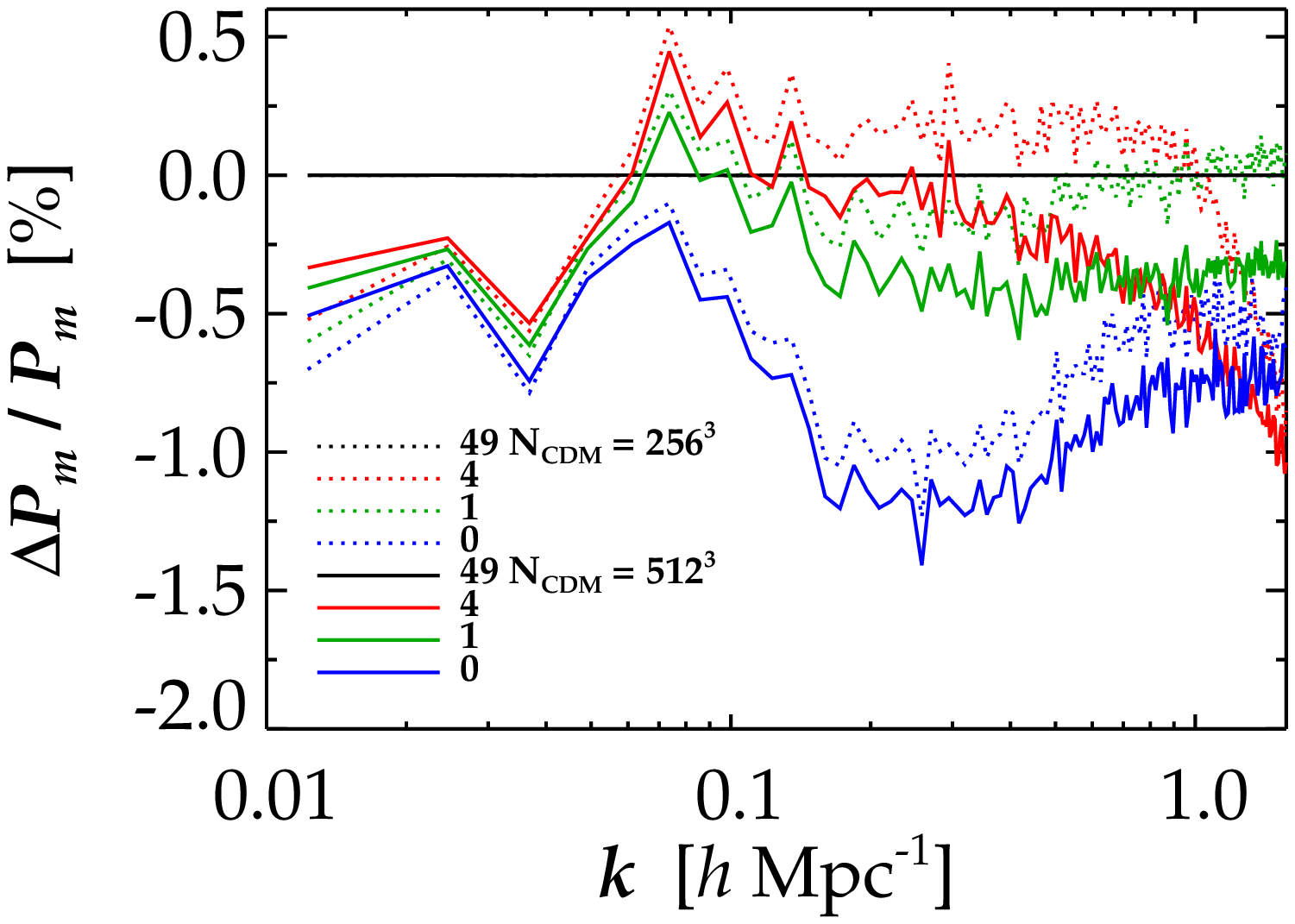}
   \end{minipage}
   \caption{Left: The non-linear neutrino correction as a function of the simulation volume and redshift. Dotted lines is for a $256 \, h^{-1} {\rm Mpc}$ box ($C_4/C_5$) and solid lines for our standard $512 \, h^{-1} {\rm Mpc}$ box ($C_1/C_3$). Right: Convergence as a function of the number of CDM $N$-body particles at different redshifts. Dotted lines is for $256^3$ CDM particles ($C_6/C_7$) and solid lines for $512^3$ CDM particles ($C_1/C_2$). In both figures $\sum m_\nu = 0.6 \, {\rm eV}$.}
   \label{fig:converge_test}
\end{figure}

\begin{figure}[t]
   \noindent
   \begin{minipage}{0.49\linewidth}
      \hspace*{-1.4cm}\includegraphics[width=1.25\linewidth]{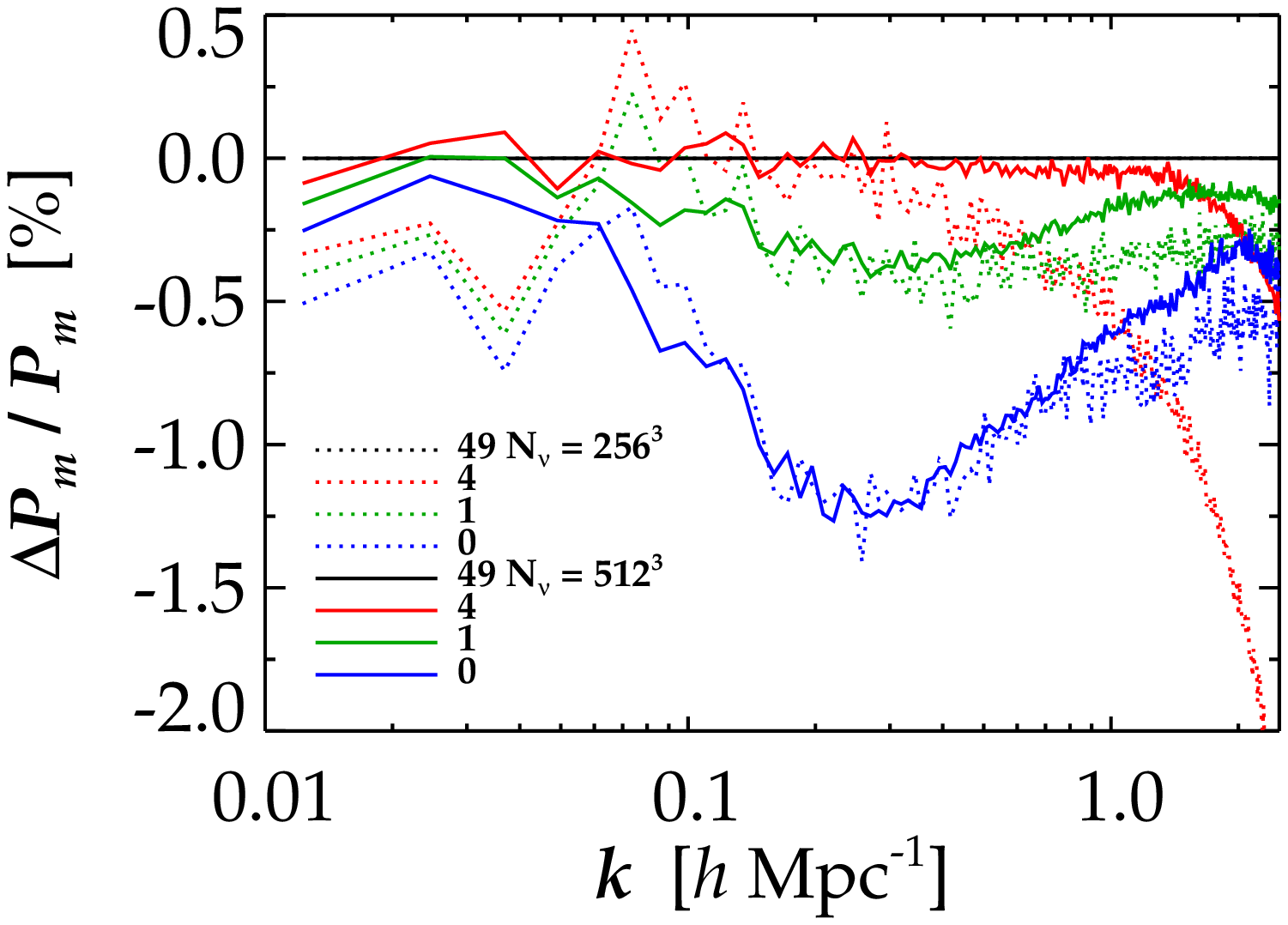}
   \end{minipage}
   \begin{minipage}{0.49\linewidth}
      \hspace*{-0.3cm}\includegraphics[width=1.25\linewidth]{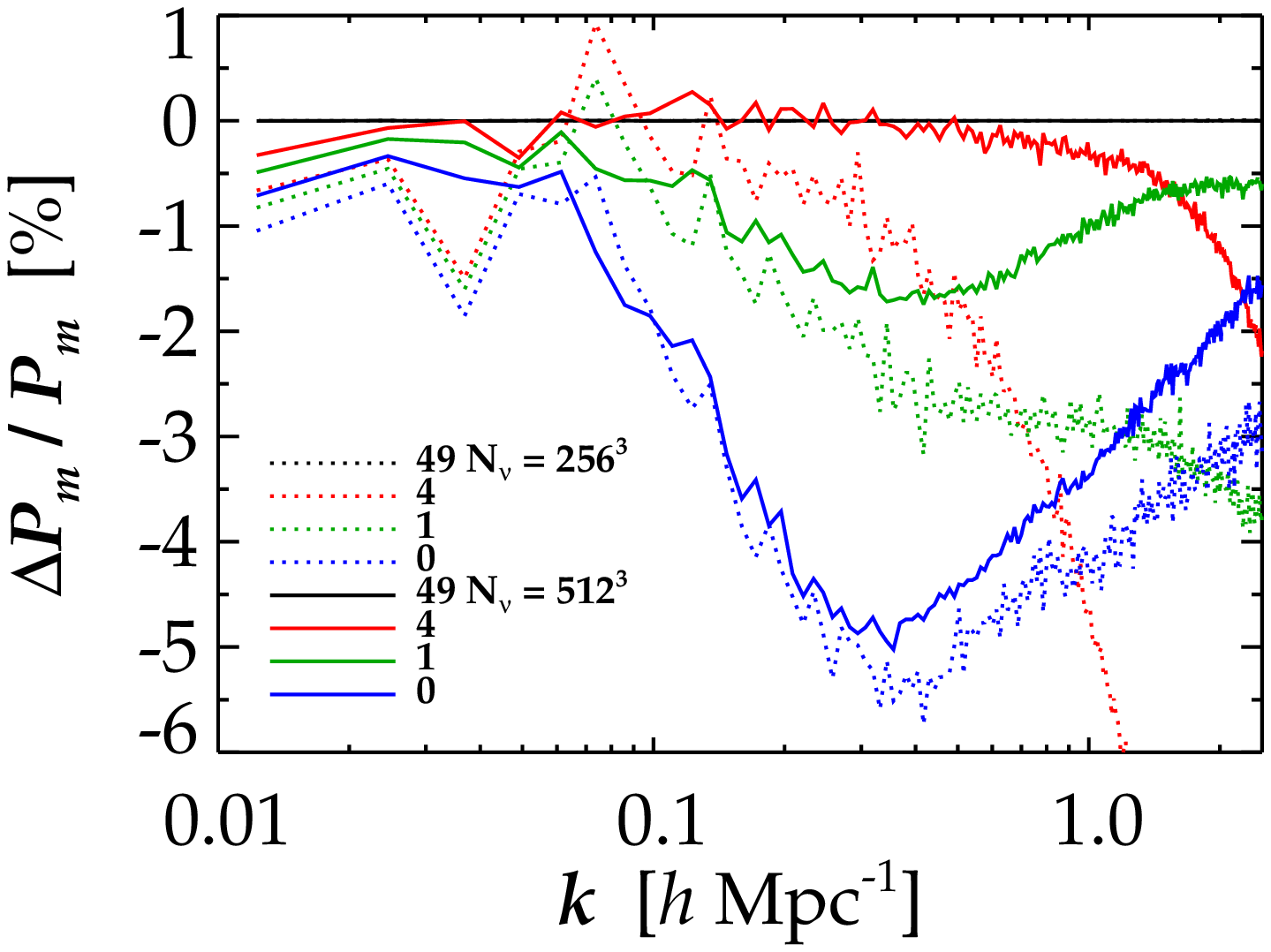}
   \end{minipage}
   \caption{Convergence as a function of neutrino $N$-body particles at various redshifts. Left: $\sum m_\nu = 0.6 \, {\rm eV}$, dotted lines is for $256^3$ neutrino particles ($C_1/C_2$) and solid lines for $512^3$ neutrino particles ($C_1/C_3$). Right: $\sum m_\nu = 1.2 \, {\rm eV}$, dotted lines is for $256^3$ neutrino particles ($D_1/D_2$) and solid lines for $512^3$ neutrino particles ($D_1/D_3$).}
   \label{fig:nu_test}
\end{figure}

\subsection{Box Size and Particle Shot Noise}
The finite size of the simulation volume as well as a limited number of $N$-body particles can significantly affect the simulated power spectrum (see \cite{Heitmann08} for a recent analyses). We have chosen a box size of $512 \, h^{-1} {\rm Mpc}$ from the considerations that a significant number density of neutrino $N$-body particles was needed to suppress the thermal velocity white noise term on small scales. Increasing the box size by a factor of 2 would demand a factor of 8 more neutrino $N$-body particles, stretching our computational resources, especially CPU time, to the limits.

But, as shown in the left panel of Fig.~\ref{fig:converge_test}, our chosen box size is sufficiently large for testing the neutrino grid method. The figure shows the non-linear correction from our highest resolution runs with $\sum m_\nu = 0.6 \, {\rm eV}$ ($C_1/C_3$), compared against a simulation where the box size, number of CDM and neutrino particles as well as the PM grid have been scaled down by a factor of 2 per dimension ($C_4/C_5$). Note that the two sets of simulations have been started from different initial random numbers. The difference in the calculated non-linear correction is at the 0.2\% level at most. Since we expect rapid convergence as the box size is increased, the chosen box size of $512 \, h^{-1} {\rm Mpc}$ is sufficient.

Including enough CDM particles is important for calculating the gravitational potential accurately since the neutrino $N$-body particles move on this background. The difference between including $256^3$ ($C_6/C_7$) and $512^3$ ($C_1/C_2$) CDM particles on the calculation of the non-linear neutrino correction is seen in the right panel of Fig.~\ref{fig:converge_test}. At higher redshift the difference is largest at smaller scales, but the two sets of simulations begin to converge at low redshift. Note that the lower resolution CDM simulation predicts a slighly smaller non-linear neutrino correction at the 0.2\% level. Again, expecting the size of the non-linear neutrino correction term to converge rapidly as the number of CDM particles are increased, our choose of $512^3$ CDM particles is sufficient. Also note that the neutrino particle and grid methods approach each other at the largest scales as more CDM particles are included and that the differences between the two CDM resolutions mainly build up at high redshift, where the particle distributions are most homogeneous.

Fig.~\ref{fig:nu_test} shows convergence as a function of $256^3$ or $512^3$ neutrino $N$-body particles in the $\sum m_\nu = 0.6 \, {\rm eV}$ and $1.2 \, {\rm eV}$ cases. Focusing on the lower mass neutrinos it can be seen that the discrepancy between the $256^3$ and $512^3$ simulations is largest at the smallest and largest scales simulated. At the smallest scales since here the neutrino white noise term contributes substantially to the total matter power spectrum, and at the largest scales since a noise term generated at high redshift is added to a real significant gravitational signal and therefore persists today. We conclude that in the range $k \simeq 0.1 - 1 \, h {\rm Mpc}^{-1}$ where non-linear correction terms to the neutrino component are important the difference between the neutrino grid and particle methods has converged as a function of the number of neutrino $N$-body particles.

In the $\sum m_\nu = 1.2 \, {\rm eV}$ case the noise term is even more pronounced at small scales, $k \gtrsim 0.3 \, h {\rm Mpc}^{-1}$, since now $\Omega_\nu$ is larger even though the thermal velocity is smaller. As demonstrated in \cite{Brandbyge1} the noise term decreases rapidly as the number of neutrino $N$-body particles is increased, therefore we expect the difference between the neutrino grid and particle methods to have almost converged for $512^3$ neutrino $N$-body particles.

Since the maximum non-linear neutrino correction in the $1.2 \, {\rm eV}$ case has almost converged for $512^3$ neutrino $N$-body particles, and it has converged for $0.6 \, {\rm eV}$ neutrinos, scaling to the $0.3 \, {\rm eV}$ case we see that the 0.25\% correction here should also have converged.

With respect to the neutrino grid we (usually) have a Nyquist frequency of $\pi \, h {\rm Mpc}^{-1}$. This is an oversampling of the neutrino perturbations, so that linear neutrino modes are not missing at scales where they contribute. We chose a cut-off in the neutrino grid modes at the grid Nyquist frequency.

\subsection{Time and Force Resolution}

To get the neutrino Fourier modes at a specified redshift we have interpolated linearly in the scale factor between the TFs in the library. These TFs are themselves equally spaced in the scale factor. To test the accuracy of this method we have run simulations with $\sum m_\nu = 0.6 \, {\rm eV}$ where the number of TFs in the library were decreased from 500 by a factor of 5. On scales where the neutrinos contribute to the matter power spectrum, the difference found was below 0.05\% for all redshifts. Also note that the non-linear neutrino correction term found is not due to an inaccurate sampling of the TFs, since such an error would manifest itself especially on the largest scales where the neutrino component contributes the most.

We also tested our choices for the \textsc{gadget}-2 parameters {\small{ErrTolIntAccuracy}} $ = 0.025$ and {\small{MaxSizeTimeStep}} $= 0.03$ by decreasing them by a factor of 4 and 6, respectively.  At $k < 1 \, h {\rm Mpc}^{-1}$ the difference is at most 0.1 \% in each case, so regarding these two parameters the absolute power spectrum is evolved accurately in the region where we have the non-linear neutrino correction. We also ran the code with double precision and found an absolute error below $0.1\%$, which only manifested itself on small scales.

Since the neutrino particle component mainly contributes to the gravitational potential in the region coverged by the PM grid, testing for convergence of the long range force is important, so as not to introduce errors when comparing the neutrino grid and particle methods. Therefore, we have run two simulations with a PM grid size of $1024$ (models $C_8$ and $C_9$). These simulations gave an almost identical calculation of the non-linear neutrino correction term compared to the $N_{\rm PM} = 512$ case.

In the standard TreePM approach the long range force is Gaussian filtered. Since we do not filter the neutrino grid modes, these modes contribute at smaller scales than do the neutrino particle long range modes. Since the PM approach breaks down at small scales, it is necessary to test how much this lack of filtering contributes. We have done this by running our 3 neutrino grid simulations (models $B_1$, $C_1$ and $D_1$) with the Gaussian filtering included.

In the $\sum m_\nu = 1.2 \, {\rm eV}$ case the non-linear neutrino correction term increases by only $\simeq 0.25\%$ with the smoothing turned on. For the lower neutrino masses the effect is negligible ($\lesssim 0.05\%$). Therefore, including the neutrino grid modes via the PM approach without a Gaussian filtering is well justified, since the neutrinos do not contribute substantially to the gravitational potential on scales where the PM method breaks down. Or equivalently, it is extremely well justified to neglect the linear neutrino component in the Tree part (for our chosen $N$-body specific parameters, i.e. $R_{\rm BOX} = 512 \, h^{-1}{\rm Mpc}$, $N_{\rm PM}=512$, $r_{\rm s} = 1.25$ and $r_{cut} = 4.5$).

\subsection{The Extent in Fourier Space of the Neutrino Grid}
\begin{figure}
   \noindent
   \begin{minipage}{0.49\linewidth}
      \hspace*{-0.8cm}\includegraphics[width=1.15\linewidth]{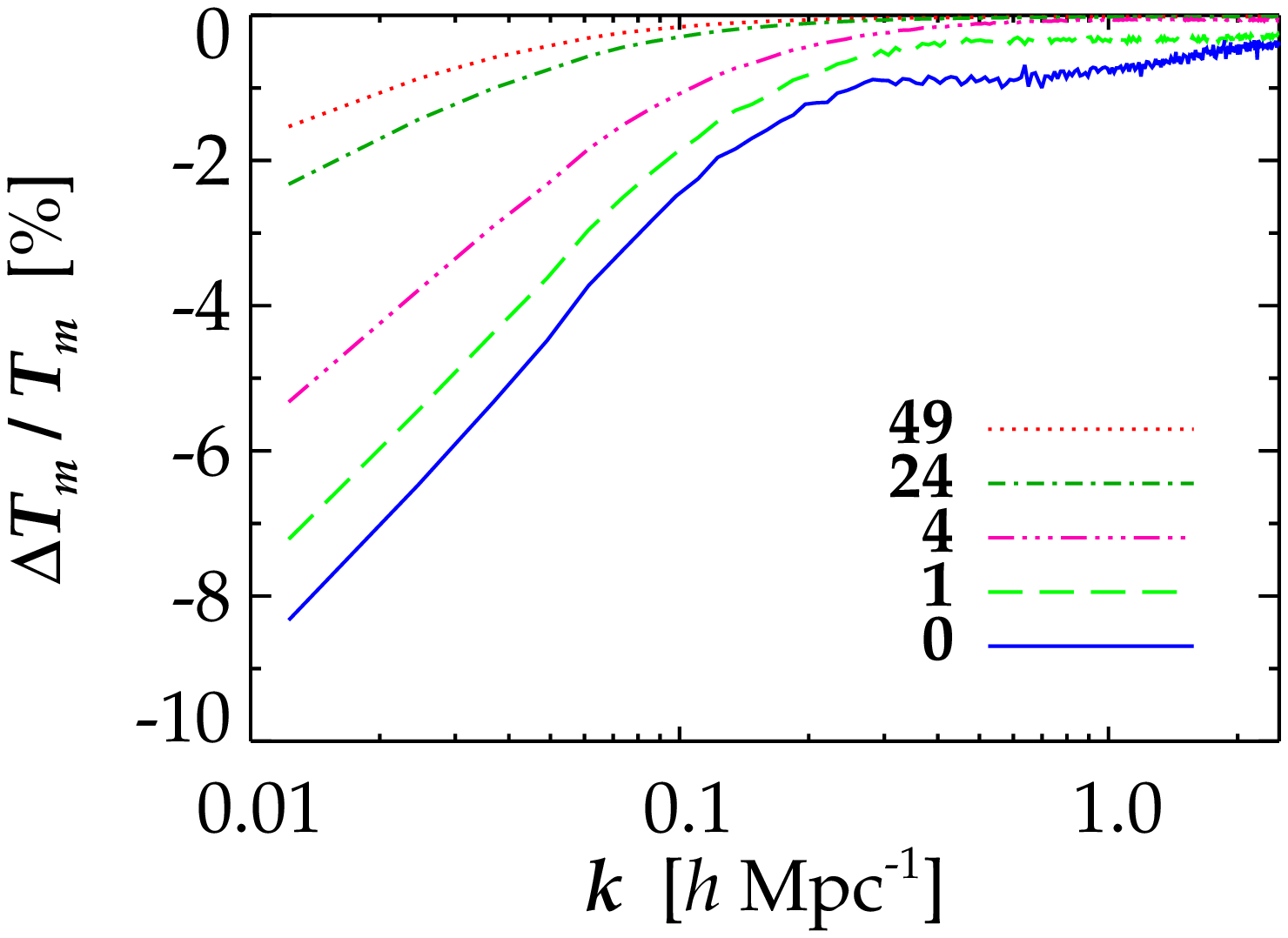}
   \end{minipage}
   \begin{minipage}{0.49\linewidth}
      \hspace*{-0.3cm}\includegraphics[width=1.15\linewidth]{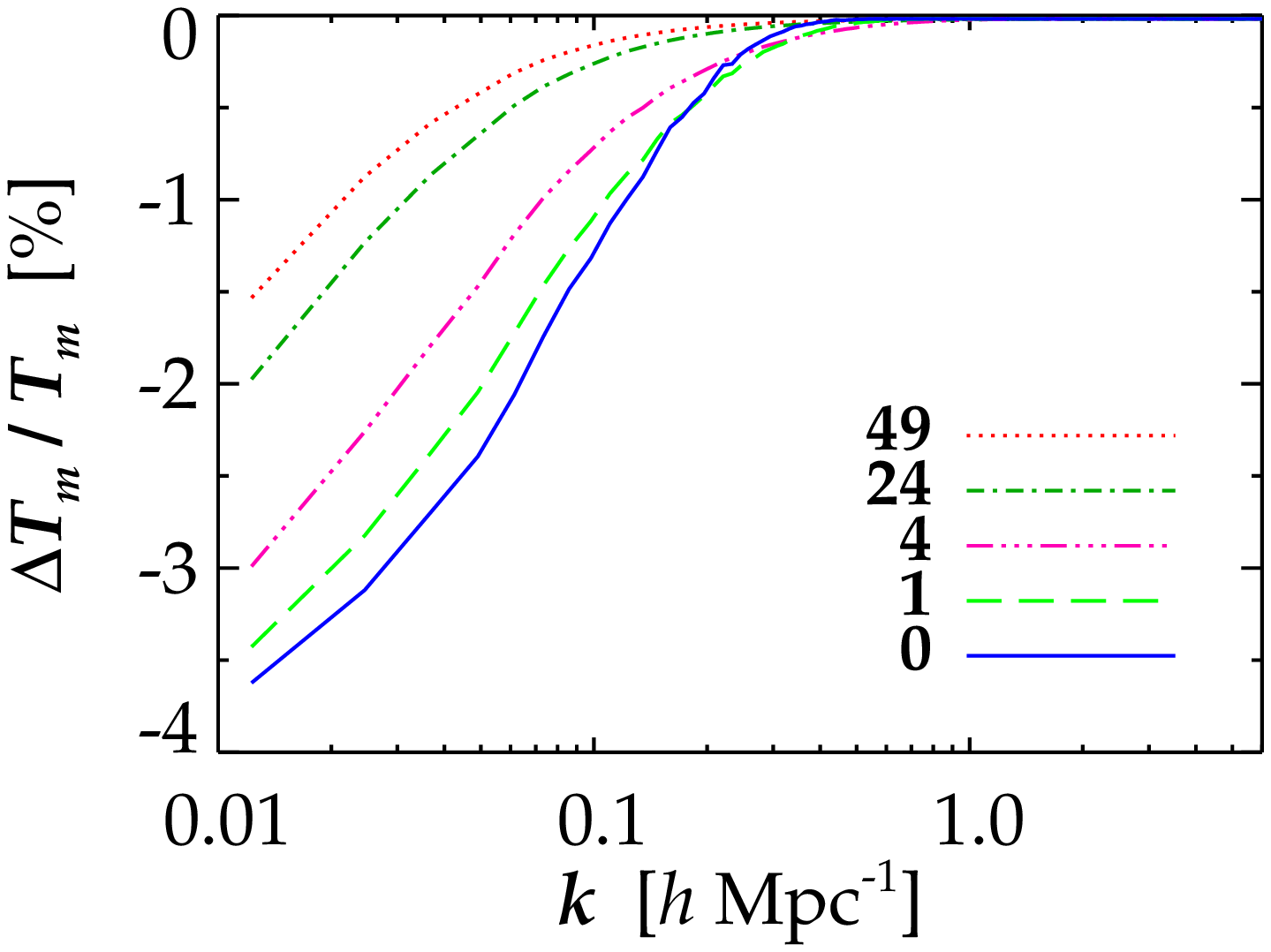}
   \end{minipage}
   \caption{Left: The effect on the square root of the matter power spectrum (an effective non-linear matter TF, $T_{\rm m}$) of neglecting the perturbed neutrino component, $\sum m_\nu = 0.6 \, {\rm eV}$, in the $N$-body simulation when it is started at $z = 49$ ($C_{10} / C_1$). Right: The effect on $T_{\rm m}$ of neglecting the neutrino component, $\sum m_\nu = 0.6 \, {\rm eV}$, only when the power spectrum is calculated at a given redshift (model $C_1$).}
   \label{fig:no_nus}
\end{figure}

In order not to oversample the neutrinos in the grid approach it is usefull to find a maximum wavenumber beyond which the neutrinos do not contribute to the formation of structure. In the left panel of Fig.~\ref{fig:no_nus} we show the effect of neglecting the $\sum m_\nu = 0.6 \, {\rm eV}$ neutrino grid from $z = 49$ onwards, i.e. throughout the whole $N$-body simulation. The neutrino component is still included in the calculation of the background evolution. The effect is of course pronounced at large scales, but even a lack of power is visible at small scales. The neutrino modes corresponding to these small scales do not contribute to this difference, as can be seen from the right panel of Fig.~\ref{fig:no_nus}. Instead, it is power missing from a lack of mode-coupling between the small-scale CDM modes and large-scale CDM plus neutrino modes. Note that we have shown the difference in the square root of the matter power spectrum, since it is this quantity that is related to the gravitational potential calculated in Fourier space in \textsc{gadget}-2.

To find a maximum wavenumber beyond which the neutrino grid does not contribute to the gravitational potential, the difference in the square root of the matter power spectrum with and without the neutrino grid is shown in the right panel of Fig.~\ref{fig:no_nus} for $\sum m_\nu = 0.6 \, {\rm eV}$. The grid is only taken out when the power spectrum is calculated and the neutrino modes are therefore included in the evolution of the matter perturbations. The wavenumber at which the difference goes to zero can then be used to find the required extent of the neutrino grid in Fourier space. For $\sum m_\nu = 0.6 \, {\rm eV}$ a conservative value for this maximum wavenumber is $k \simeq 1 \, h \, {\rm Mpc}^{-1}$. Notice that this maximum wavenumber is not found by considering the power spectra today, but at a redshift of $\simeq 4$. For $\sum m_\nu = 0.3 \, {\rm eV}$ the maximum wavenumber is $\simeq 0.5 \, h \, {\rm Mpc}^{-1}$. We caution that these maximum wavenumbers depend on $\Omega_{\rm m}$.

\section{Discussion and Conclusions}\label{DISCUSSION}

We have presented a new method for implementing neutrinos in $N$-body simulations which works extremely well for a total neutrino mass below $\sim 0.5$ eV. For such masses the difference in the matter power spectrum compared with simulations where neutrinos are treated as particles is always below 1\% on all scales. The precision is better for smaller masses, with the difference scaling roughly as $m_\nu^2$, and the gain in computational speed compared to representing the neutrinos as $N$-body particles is very large (scaling roughly as $m_\nu^{-1}$). For all masses this is the only computationally feasible way to include neutrinos in simulations at the required level of precision, especially for high $N$-body starting redshifts.

We also note that the method presented here will work for any type of energy density which has almost linear perturbations.

Finally we caution that although the method is very accurate for calculating all observables related to matter fluctuations, i.e.\ the power spectrum, halo mass functions etc, it is not accurate at the 1\% level in describing the neutrino fluctuations alone. For predicting quantities such as the local density of relic neutrinos additional steps should be taken.
One possibility is to use the 1-particle Boltzmann technique \cite{Singh:2002de,Ringwald:2004np}. In its simplest form, however, this method represents the neutrino component only in an approximate way.
Another path is to solve with the method presented here until $z$ becomes sufficiently low (in practise $z \lesssim 4$) that the noise from the neutrino thermal velocity can be kept under control. At this point the grid simulation can be used as the initial condition for a simulation with neutrinos treated as particles.

\section*{Acknowledgements}
We acknowledge computing resources from the Danish Center for
Scientific Computing (DCSC). We thank Troels Haugb{\o}lle, Julien Lesgourgues and Bjarne Thomsen for discussions.



\section*{References} 

\end{document}